\documentclass[a4paper,10pt]{article}
\usepackage{amsfonts}
\usepackage{amsmath}
\usepackage{amssymb,youngtab}
\usepackage{enumerate}
\usepackage{amsthm}
\usepackage[pdftex]{graphicx}
\usepackage{authblk}

\title{Invariants of hyperbolic Partial Differential Operators}
\author[1]{Chris Athorne}
\author[1,2]{Halis Yilmaz}
\affil[1]{School of Mathematics \& Statistics, University of Glasgow, Glasgow G12 8QW, UK}
\affil[2]{Department of Mathematics, University of Dicle, 21280 Diyarbakir, Turkey}
\date{}

\newtheorem{mydef}{Definition}[section]
\newtheorem{thm}[mydef]{Theorem}
\newtheorem{lem}[mydef]{Lemma}

\begin{document}

\maketitle

\begin{abstract}
We present a construction of a large class of Laplace invariants for linear hyperbolic partial differential operators of fairly general form and arbitrary order.
\end{abstract}

\section{Introduction}
The first examples of invariants for linear, hyperbolic partial differential operators are the Laplace invariants \cite{D}. These are defined for second order operators of the form
\[L=\partial_x\partial_y+a\partial_x+b\partial_y+c\]
and are differential functions of the coefficients, $a,b$ and $c$ which remain unchanged under conjugation by arbitrary functions of $x$ and $y:$
\[L\mapsto L^g=g^{-1}Lg.\]
They are
\begin{eqnarray}
h&=&c-ab-a,_x\nonumber\\
&=&L-(\partial_x+b)(\partial_y+a)\nonumber\\
k&=&c-ab-b,_y\nonumber\\
&=&L-(\partial_y+a)(\partial_x+b)\nonumber
\end{eqnarray}
so they may be considered obstructions to factorization in the differential ring of the coefficients of $L.$

These invariants are complete in the sense that any pair of operators $L$ and $L'$ having the same invariants are neceassrily related through conjugation by some $g.$

These invariants were important in an early theory of integrability \cite{D} because under a further class of differential transformations over the ring of coefficients (Laplace Transformations) the invariants \emph{do} change so that after a finite number of such transformations one or the other of the invariants may vanish and one has a factorized, and hence soluble, situation. Since the Laplace transformations are invertible one can solve the original equation. There is a link with modern integrability theory here: the equations relating the transformation properties of the invariants are now known as the two-dimensional Toda field theory equations \cite{T,W}.

They are of importance too in the theory of first order nonlinear systems because the classical Riemann invariants for such systems (e.g. of hydrodynamic type) satisfy second order, linear partial differential equations. Hence the link with Laplace invariants allow Laplace transformations to be lifted up to transformations of the non-linear systems \cite{F,GZ}. The role of the invariants in the study of linearizable partial differential equations of Liouville type is summarised in \cite{ZS}.

Generalizations of the above story exist. In the first place invariants for non-hyperbolic operators were discussed quite early ( see references in \cite{D}) and wider classes of differential transformation (Moutard \cite{Md}, Darboux \cite{D}, Goursat \cite{G}) were constructed for restricted classes of invariants.

In geometry Laplace invariants arise \cite{D} because the generic immersion of a surface into $\mathbb R^3$ is described, in \emph{conjugate coordinates} by a vector whose mixed second derivative lies in the tangent plane to the immersed surface, hence satisfying a second order, linear hyperbolic partial differential equation. Laplace transformations then corrrespond to transformations between surfaces with accompanying conjugate coordinates. In particular the case of a vanishing invariant describes a degenerate surface, namely a curve \cite{Kam}. Generalization to higher dimensional manifolds is dealt with in \cite{KamT}.

A formulation of Laplace invariants using the Cartan method of moving frames is given in \cite{SE}.

Modern work in the specific area of the construction of Laplace type invariants is for higher order operators \cite{A,AY,S} and in factorization issues \cite{Ts1,Ts2,SW}. In particular the question of invariants for operators of orders three and four in small dimensions are dealt with in \cite{BKar,Dz,JA,Kar,MM,S} and in four dimensions with leading term $\partial_1\partial_2\partial_3\partial_4$ in \cite{M}.

The objective of this paper is to construct such invariants for a wide class of such operators of all orders.

We start with a given index set $I$ corresponding to the $n$ labels of independent variables, $x_1,x_2,\ldots,x_n,$ and with an operator $L_I$ with leading term $\prod_{i\in I}\partial_i.$ We construct, using the coefficients of $L_I$ in a symmetric way, families of lower order operators $L_J$ for $J\subseteq I$ with leading term $\prod_{j\in J}\partial_j$ and form noncommutative polynomials in these $L_J.$ The leading order coefficient of any such polynomial is necessarily an invariant. In particular any such polynomial which is actually a function (differential operator of order zero) is necessarily an invariant and we show that, in fact, every invariant in the differential ring of the coefficients of $L_I$ arises in this way. Our goal is therefore to describe zero order operators in the ring generated by the $L_J$ for $J\subseteq I.$

Any such operator commutes with an arbitrary function, $\theta,$ which allows us to define the kernel of a map $\Theta$ from polynomials in the $L_J$ with coefficients in $\mathbb Q$ to polynomials in the $L_J$ whose coefficients are linear in derivatives of $\theta.$ We can relate $\Theta$ to a universal operator, $\delta,$ on ordered partitions of $m\leq n$ which we represent using box diagrams. $\delta$ acts as a derivative on these diagrams. An invariant is constructed from a polynomial in diagrams via a symmetric sum over index sets of order $m.$ We show by a counting argument that we are able to describe the kernel of $\delta$ completely and hence construct all invariants for $L_I$ which are symmetric in indices. There are $2^n-(n+1)$ such fundamental symmetric invariants for $L_I.$ In particular we present a generating set which has a unique ``new'' invariant at each order, all others being created by operations such as multiplication and derivation from lower order invariants.

Finally we write down, in a clear but condensed notation, the set of symmetric fundamental invariants of orders up to six.

\section{The operators}
Let $I$ be a fixed set of $n$ labels, for convenience, $I=\{1,2,\ldots n\}.$ By $J,$ $K,$ $M$ etc. we denote subsets of $I.$ If $J\subseteq I$ then $|J|$ is the cardinality of $J.$ 

We are concerned with partial differential operators of the form
\begin{equation}\label{defLJ}
L_J=\sum_{p=0}^{|J|}\left(\sum_{K\subseteq J,\,|K|=p}a_{J\backslash K}\partial_K\right)
\end{equation}
The subset subscripts on $a$ and $\partial$ indicate (by a slight abuse of notation) totally symmetric multi-indices. Thus 
\[\partial_K=\prod_{i\in K}\partial_i\]
We take $a_{\emptyset}=1.$ The $a_K$ are functions of independent variables $x_1,x_2,\ldots x_n$ and the $\partial_i$ act as derivations on the polynomial ring in the $a_K$ and their derivatives, $a_K,_M$ over $\mathbb Q.$

For example, if $n=3$ we have the following seven operators of non zero order
\begin{eqnarray}
L_i&=&\partial_i+a_i,\quad i\in \{1,2,3\}\nonumber\\
L_{ij}&=&\partial_{ij}+a_i\partial_j+a_j\partial_i+a_{ij},\quad  i,j \in \{1,2,3\}\nonumber\\
L_{ijk}&=&\partial_{ijk}+a_i\partial_{jk}+a_j\partial_{ki}+a_k\partial_{ij}\nonumber\\
&&+a_{ij}\partial_k+a_{jk}\partial_i+a_{ki}\partial_j+a_{ijk},\quad i,j,k\in\{1,2,3\}\nonumber
\end{eqnarray}

Each operator is totally symmetric in its indices. So, for example, there are three second order operators: $L_{12},L_{23}$ and $L_{31};$ and only one third order: $L_{123}.$ Any polynomial in the $a_K$ and their derivatives is a zero order differential operator. 

The $a_i,$ $a_{ij}$ etc. occuring in each of the $L_J$ are the same objects, i.e. they do not depend upon the degree of the operator.

\subsection{Invariants}
The invariants in which we are interested are related to transformations of the form
\begin{equation}
L_J\mapsto L^g_J=g^{-1}L_Jg,
\end{equation}
$g$ being an arbitrary function of the independent variables, which preserve the form of $L_J$ but alter the coefficients $a_{J\backslash K}:$
\begin{equation}
a_{J\backslash K}\mapsto a^g_{J\backslash K}\nonumber.
\end{equation}
For example:
\begin{eqnarray}
L^g_i&=&\partial_i+a^g_i\nonumber\\
L^g_{ij}&=&\partial_{ij}+a^g_i\partial_j+a^g_j\partial_i+a^g_{ij}\nonumber
\end{eqnarray}
where
\begin{eqnarray}\label{gt2}
a^g_i&=&a_i+g^{-1}g,_i\nonumber\\
a^g_{ij}&=&a_{ij}+g^{-1}g,_ja_i+g^{-1}g,_ia_j+g^{-1}g,_{ij}.
\end{eqnarray}
We use the comma notation to indicate derivatives of functions.

Note that the transformation law for the $a_{J\backslash K}$ depends only on the set of indices $J\backslash K$ regardless of the $L_J$ opeartor to which it belongs. To see this note that
\begin{eqnarray}
L^g_J&=&\sum_{K\subseteq J}a_{J\backslash K}g^{-1}\partial_Kg\nonumber\\
&=&\sum_{K\subseteq J}a_{J\backslash K}\left(\sum_{M\subseteq K}g^{-1}g,_{K\backslash M}\partial_M\right)\nonumber
\end{eqnarray}
so that 
\[a^g_{J'}=\sum_{K'\subseteq J'}a_{J'\backslash K'}g^{-1}g,_{K'}\]
for any subset $J'$ of $I.$ Thus transformations do not interfere with our requirement that the same set of coefficients occurs in operators of all degrees.

An invariant, $\mathfrak I,$ is then any function of the set of $a$'s and their derivatives which takes the same value under any such transformation:
\begin{equation}
{\mathfrak I}(\{a_J\, and\,derivatives|J\subseteq I\})={\mathfrak I}(\{a^g_J\, and\,derivatives|J\subseteq I\})
\end{equation}

In the case of (\ref{gt2}) above, invariants are easily seen to be
\begin{eqnarray}
[i,j]&=&a_j,_i-a_i,_j\nonumber\\
(i,j)&=&2a_{ij}+2a_ia_j-a_i,_j-a_j,_i\nonumber
\end{eqnarray}
using a notation introduced in an earlier paper.

A crucial observation is that we can express invariants as differential operators of order zero constructed out of the $L_J$ for $J\subseteq I.$ In the case above
\begin{eqnarray}
[i,j]&=&L_iL_j-L_jL_i\nonumber\\
(i,j)&=&2L_{ij}-L_iL_j-L_jL_i\nonumber
\end{eqnarray}

Clearly any polynomial in the $L_J$ which is, by dint of cancellations, a zeroth order differential operator, is an invariant because conjugation by $g$ willl leave it unaffected. Its derivatives will also be invariants but these can be expressed as (noncommutative) polynomials too since
\[{\mathfrak I},_k=[L_k-a_k,{\mathfrak I}]=L_k{\mathfrak I}-{\mathfrak I}L_k.\]

We have the stronger statement:

\begin{thm}
The invariants of the operator $L_I$ are (noncommutative) polynomials in the $L_J$ for $J\subseteq I.$
\end{thm}

\emph{Proof.} 
It is clear, by linear algebra, that we can express all the $a_J$ coefficients of $L_I$ in terms of the $L_K$ for $K\subseteq J$ and $\partial_{J\backslash K}$ in the following way:
\begin{equation}\label{relation}
a_J=\sum_{p=0}^{|J|}\left(\sum_{K\subseteq J,\,|K|=p}(-1)^{|K|}L_{J\backslash K}\partial_K\right)
\end{equation}

Any polynomial $\mathfrak I$  in the $a_J$ and their derivatives can therefore be expressed as a  polynomial in the $L_J$ and the $\partial_k:$ 
\[\mathfrak I(a_J,\ldots|J\subseteq I)=F(L_J,\partial_1,\ldots,\partial_n|J\subseteq I).\] 

The invariance condition is \[\mathfrak I(a_J,\ldots|J\subseteq I)=\mathfrak I(a^g_J,\ldots|J\subseteq I)\] and since the relation (\ref{relation})  between the $a_J$ and the $L_J$ holds equally for the $a^g_J$ and the $L^g_J$ we can write
\[F(L_J,\partial_1,\ldots,\partial_n|J\subseteq I)=F(L^g_J,\partial_1,\ldots,\partial_n|J\subseteq I).\] 

Under conjugation by $g$ since $F$ is a zeroth order differential operator we have

\[F(L_J,\partial_1,\ldots,\partial_n|J\subseteq I)=F(L^g_J,\partial^g_1,\ldots,\partial^g_n|J\subseteq I)\]
where $\partial^g_i=\partial_i+g^{-1}g,_i.$ There last two statementss allow us to write
\[F(L_J,\partial_1,\ldots,\partial_n|J\subseteq I)=F(L_J,\partial^h_1,\ldots,\partial^h_n|J\subseteq I)\]
for arbitrary $h.$

This implies that $F$ is a function of the $L_J$ only. Otherwise we could make choices of $h$ (e.g. $h=e^{x_i}$) which, since both sides are zeroth order differential operators, would describe differential relations between the $a_J.$ The coefficients of $L_J$ have, by assumption, no such relations. Hence the result. $\blacksquare$

\subsection{Degrees}
We define several different degrees.

The \emph{differential} degree of a term $a_J,_K\partial_M$ is $|M|,$ the \emph{coefficient} degree is $|J|+|K|$ and the \emph{total} degree is $|J|+|K|+|M|.$ Clearly the total degree of any of the $L_J$ is just $J.$ We will say that the monomial $L_{J_1}L_{J_2}\ldots L_{J_p}$ has $L$-degree $|J_1|+|J_2|+\ldots |J_p|$ which is the same as its total degree. We extend this notion to polynomials in the usual way. Of course the differential degree of a polynomial in the $L$'s may be smaller than its total degree.

Thus the two invariants listed above are of differential degree zero but total degree and $L$-degree two. Any non-zero $L$-degree object of differential degree zero will be an invariant. We note the following result \cite{C}.

\begin{thm}\label{commutatordegree}
The differential degree of $L_JL_K-L_KL_J$ is $|J|+|K|-2$.
\end{thm}

As a generalisation of our earlier characterisation of invariants we note that:

\begin{thm}\label{highest}
The coefficient of the term of highest differential degree in any polynomial in the $L_J,\, J\subseteq I,$ is an invariant.
\end{thm}

\emph{Proof.} The term of highest differential degree will be of the form $f\prod_{i=1}^n\partial_i^{n_i}$ where $f$ is a function of the $a_J$ and their derivatives. Under conjugation $\partial_i\mapsto \partial_i+g^{-1}g,_i$ which leaves the highest degree term unaltered. $\blacksquare$

\subsection{The $\Theta$ map.}
A differential operator $L$ of order zero is characterised by the property that it commutes with any function. Hence for some arbitrary function $\theta$ we define the map
\begin{equation}
\Theta: L\mapsto \Theta(L)=[L,\theta]
\end{equation}
and we seek to describe the kernel of $\Theta$ in the spaces of polynomials of each $L$-degree.

As will be seen by studying examples we should think of $\Theta$ as a map from the vector space of polynomials  of $L$-degree equal to $N$ over constant coefficients to the space  polynomials of $L$-degree strictly less than $N$ with coefficients linear in derivatives of $\theta.$ For example

\begin{eqnarray}
\Theta(L_i)&=&\theta,_i\nonumber\\
\Theta(L_{ij})&=&\theta,_{ij}+\theta,_iL_j+\theta,_jL_i\nonumber\\
\Theta(L_{ijk})&=&\theta,_{ijk}+\theta,_{ij}L_k+\theta,_{jk}L_i+\theta,_{ki}L_j+\nonumber\\
&&\theta,_iL_{jk}+\theta,_jL_{ki}+\theta,_kL_{ij}\nonumber
\end{eqnarray}

In general,

\begin{thm}\label{Theta}
\[\Theta(L_J)=\sum_{\emptyset\neq K\subseteq J}\theta,_{ K}L_{J\backslash K}.\]
\end{thm}

\emph{Proof.} We need only to compare the coefficients of the (multi-)derivatives of $\theta$ on each side of this equation and we can do this by making a special choice for $\theta.$ So let $\theta=\prod_{i\in M}x_i$ where $M\subseteq J.$ Denote this function by $x_M.$

For $|M|=1$, $\theta=x_i$ and so $\theta,_K,$ (for $|K|\geq 1$) is vanishing unless $K=\{i\}.$ On the other hand
\[
[\partial_K,x_i]=
\left\{
\begin{array}{cc}
\partial_{K\backslash\{i\}} & i\in K\\
0 & i\notin K
\end{array}
\right.
\]

Then (from (\ref{defLJ})) the left hand side of the statement of the theorem is
\[\Theta(L_J)=\sum_{i\in K\subseteq J}a_{J\backslash K}\partial_{K\backslash\{i\}}=L_{J\backslash\{i\}}\]
and the right hand side is
\[\sum_{K=\{i\}}\theta,_{ K}L_{J\backslash K}=L_{J\backslash\{i\}}.\]

Hence the result is established for $|M|=1.$

Assume the result true for $|M|=m$ and let $M'=M\cup\{i\}$ for $i\notin M.$  Then $\theta=x_{M'}=x_Mx_i.$ 

The left hand side of the equality in the statement of the theorem is
\begin{eqnarray}\label{lhs}
[L_J,x_Mx_i]&=&[L_J,x_M]x_i+x_M[L_J,x_i]\nonumber\\
&=&\sum_{\emptyset\neq K\subseteq J}x_M,_KL_{J\backslash K}x_i+x_ML_{J\backslash\{i\}}
\end{eqnarray}
by our inductive assumption and the case $|M|=1.$

The right hand side of the equality in the statement of the theorem is

\begin{eqnarray}\label{rhs}
\sum_{\emptyset\neq K\subseteq J}(x_Mx_i),_KL_{J\backslash K} & = & \sum_{i\in K\subseteq J}x_M,_{K\backslash\{i\}}L_{J\backslash K}+\sum_{i\notin K\subseteq J,K\neq\emptyset}x_M,_Kx_iL_{J\backslash K}\nonumber\\
&=&\sum_{i\in K\subseteq J}x_M,_{K\backslash\{i\}}L_{J\backslash K}+\sum_{i\notin K\subseteq J,K\neq\emptyset}x_M,_KL_{J\backslash K}x_i\nonumber\\
&&-\sum_{i\notin K\subseteq J,K\neq\emptyset}x_M,_K[L_{J\backslash K},x_i]\nonumber\\
&=&\sum_{i\notin K\subseteq J,K\neq\emptyset}x_M,_KL_{J\backslash K}x_i+\nonumber\\
&&\sum_{i\in K\subseteq J}x_M,_{K\backslash\{i\}}L_{J\backslash K}-\sum_{i\notin K\subseteq J,K\neq\emptyset}x_M,_KL_{J\backslash (K\cup\{i\})}\nonumber\\
&=&\sum_{i\notin K\subseteq J,K\neq\emptyset}x_M,_KL_{J\backslash K}x_i+x_ML_{J\backslash\{i\}}\nonumber\\
&=&\sum_{\emptyset\neq K\subseteq J}x_M,_KL_{J\backslash K}x_i+x_ML_{J\backslash\{i\}}
\end{eqnarray}
since $i\notin M.$ The equality of (\ref{lhs}) and (\ref{rhs}) prove the result for all $J$ and $M$ by induction.$\blacksquare$

A more general result for the action of $\Theta$ on monomials $L_{J_1}L_{J_2}\ldots L_{J_p}$ is:

\begin{thm}\label{ThetaMonomial}
Let $J_i$ for $i=1,\ldots p$ be a set of pairwise disjoint subsets of $I.$ Then
\[
\Theta(L_{J_1}L_{J_2}\ldots L_{J_p})=\sum_{K_i\subseteq J_i,\, \bigcup_iK_i\neq\emptyset}\theta,_{K_1K_2\ldots K_p}L_{J_1\backslash K_1}L_{J_2\backslash K_2}\ldots L_{J_p\backslash K_p}
\]
\end{thm}

\emph{Proof.} This is a straightforward inductive argument on $p$ where theorem \ref{Theta} is the case $p=1.$ $\blacksquare$

\subsection{Symmetrization}
A piece of machinery is now introduced which smooths out choices of index sets and so simplifies calculations.

If $F_K$ is any object depending on the index set $K$ then
\begin{equation}
\Sigma_J(F_K)=\sum_{\phi:K\hookrightarrow J}F_{\phi(K)}
\end{equation}
the sum being over all injections of $K$ into $J.$

For example:
\begin{eqnarray}
\Sigma_{\{1,2\}}(a_{ij})&=&a_{12}+a_{21}\nonumber\\
\Sigma_{\{1,2,3\}}(a_{ij})&=&a_{12}+a_{21}+a_{13}+a_{31}+a_{23}+a_{32}\nonumber\\
&=&2(a_{12}+a_{23}+a_{31})\nonumber\\
\Sigma_{I}(a_I)&=&|I|!a_I\nonumber
\end{eqnarray}

It is important to note that this process will kill any antisymmetric parts of the polynomials in $L_J.$ For example, any combination of terms that can be written, say, $P[L_{ij},L_{kl}]Q$ will be in the kernel of $\Sigma.$ Hence in what follows we deal only with invariants which are totally symmetric in all indices.

We apply the $\Sigma$-operator to the result of theorem \ref{ThetaMonomial}. First introduce an abbreviated notation. An object $A_K$ with subscript list $K$ is to be thought of as a specific instance of an object $A_{\mathcal Y}$ where $\mathcal Y$ is a row of empty boxes of length $|Y|.$ Then distinct terms on the right hand side of the result become indistinguishable under $\Sigma.$ Consider, for example,
\begin{eqnarray}
\Theta(L_{ij}L_{klm})&=&\theta_iL_jL_{klm}+\theta_jL_iL_{klm}+\theta_kL_{ij}L_{lm}+\theta_lL_{ij}L_{km}+\theta_mL_{ij}L_{kl}\nonumber\\
&&+\theta_{ij}L_{klm}+\theta_{ik}L_jL_{lm}+\theta_{il}L_jL_{km}+\theta_{im}L_jL_{lk}\nonumber\\
&&+\theta_{jk}L_iL_{lm}+\theta_{jl}L_iL_{km}+\theta_{jm}L_iL_{lk}\nonumber\\
&&+\theta_{kl}L_{ij}L_m+\theta_{km}L_{ij}L_l+\theta_{ml}L_{ij}L_k\nonumber\\
&&+\theta_{klm}L_{ij}+\theta_{ijk}L_{lm}+\theta_{ijl}L_{km}+\theta_{ijm}L_{kl}\nonumber\\
&&+\theta_{ikl}L_jL_m+\theta_{jkl}L_iL_m+\theta_{ikm}L_jL_l+\theta_{jkm}L_iL_l\nonumber\\
&&+\theta_{ilm}L_jL_k+\theta_{jlm}L_iL_k\nonumber\\
&&+\theta_{iklm}L_j+\theta_{jklm}L_i+\theta_{ijlk}L_m+\theta_{ijkm}L_l+\theta_{ijlm}L_k\nonumber\\
&&+\theta_{ijklm}\nonumber
\end{eqnarray}

Let $K=\{i,j,k,l,m\}.$ Then
\begin{eqnarray}\label{example}
\Sigma_K\left(\Theta(L_{\tiny{\yng(2)}}L_{\tiny{\yng(3)}}\right)
&=&\Sigma_K\,\left(\theta_{\tiny{\yng(1)}}(2L_{\tiny{\yng(1)}}L_{\tiny{\yng(3)}}+3L_{\tiny{\yng(2)}}L_{\tiny{\yng(2)}})\right.\nonumber\\
&&+\theta_{\tiny{\yng(2)}}(L_{\tiny{\yng(3)}}+6L_{\tiny{\yng(1)}}L_{\tiny{\yng(2)}}+3L_{\tiny{\yng(2)}}L_{\tiny{\yng(1)}})\nonumber\\
&&+\theta_{\tiny{\yng(3)}}(4L_{\tiny{\yng(2)}}+6L_{\tiny{\yng(1)}}L_{\tiny{\yng(1)}})\nonumber\\
&&+\theta_{\tiny{\yng(4)}}5L_{\tiny{\yng(1)}}+\left.\theta_{\tiny{\yng(5)}}\right)
\end{eqnarray}

Let us denote by $\delta$ the operator that removes a box from any of the $L_{\mathcal Y}$ in any way. So
\begin{eqnarray}
\delta(L_{\tiny{\yng(2)}}L_{\tiny{\yng(3)}})&=&2L_{\tiny{\yng(1)}}L_{\tiny{\yng(3)}}+3L_{\tiny{\yng(2)}}L_{\tiny{\yng(2)}}\nonumber\\
\delta^2(L_{\tiny{\yng(2)}}L_{\tiny{\yng(3)}})&=&2L_{\tiny{\yng(3)}}+12L_{\tiny{\yng(1)}}L_{\tiny{\yng(2)}}+6L_{\tiny{\yng(2)}}L_{\tiny{\yng(1)}}\nonumber
\end{eqnarray}
and so on. 

$\delta$ is nilpotent on any finite diagram. Denote by $\theta_{\bf n}$ the symbol $\theta$ subscripted by $n$ empty boxes. Then, as operators on finite polynomials in the $L_J,$

\begin{equation}\label{exp}
\Sigma_K\Theta=\sum_{n=1}^\infty\frac{1}{n!}\theta_{\bf n}\delta^n.
\end{equation}

\begin{thm}
A sum of products of the $L_J$ with rational coefficients each monomial summand of which has the same $L$-degree and which is totally symmetric in its indices will be an invariant if and only if the corresponding sum of products of $L_{\mathcal Y}$  lies in the kernel of $\delta.$
\end{thm}

\emph{Proof.} 
Clearly if the sum of products of $L_Y$ lies in the kernel of $\delta$ then the polynomial in $L_J$ lies in the kernel of $\Theta.$ 

Suppose now there is such an invariant polynomial in the $L_J.$ It will have a corresponding polynomial in the $L_{\mathcal Y}$ which is in the kernel of the operator on the right hand side of (\ref{exp}). Since the coefficients $\theta_{\bf n}$ are all independent it must lie in the kernel of each $\delta^n$ for $n\geq 1.$ Hence the result. $\blacksquare.$

Thus, for example,
\begin{eqnarray}
\delta(L_{\tiny{\yng(3)}}-3L_{\tiny{\yng(2)}}L_{\tiny{\yng(1)}}+2L_{\tiny{\yng(1)}}L_{\tiny{\yng(1)}}L_{\tiny{\yng(1)}})&=&3L_{\tiny{\yng(2)}}\nonumber\\
&&-6L_{\tiny{\yng(1)}}L_{\tiny{\yng(1)}}-3L_{\tiny{\yng(2)}}\nonumber\\
&&+6L_{\tiny{\yng(1)}}L_{\tiny{\yng(1)}}\nonumber\\
&=&0\nonumber
\end{eqnarray}

A totally symmetric sum of insertions of any three distinct indices into the degree three polynomial on the left hand side is an invariant. Thus, using the symmetry properties of the $L_J,$
\begin{eqnarray}
\Sigma_{\{i,j,k\}}(L_{\tiny{\yng(3)}}-3L_{\tiny{\yng(2)}}L_{\tiny{\yng(1)}}+2L_{\tiny{\yng(1)}}L_{\tiny{\yng(1)}}L_{\tiny{\yng(1)}})&=&6L_{ijk}\nonumber\\
&&-6L_{ij}L_{k}-6L_{jk}L_{i}-6L_{ki}L_{j}\nonumber\\
&&+2L_iL_jL_k+2L_jL_iL_k+2L_iL_kL_j\nonumber\\
&&+2L_kL_iL_j+2L_jL_kL_i+L_kL_jL_i\nonumber\\
&=&6a_{ijk}-6a_{ij}a_k-6a_{jk}a_i-6a_{ik}a_j\nonumber\\
&&+12a_ia_ja_k-2a_{i,jk}-2a_{j,ki}-2a_{k,ij}\nonumber
\end{eqnarray}

In the final section (\ref{eggs}) we use an overbar notation to mean total symmetrization over all indices and we choose numeric labels as representatives of any set. Thus we will write the above invariant,
\begin{eqnarray}
I_{3}&=&\overline{L_{123}}-3\overline{L_{12}L_{3}}+2\overline{L_{1}L_{2}L_{3}}\nonumber\\
&=&\overline{a_{123}}-3\overline{a_{12}a_{3}}+2\overline{a_{1}a_{2}a_{3}}-\overline{a_{1,23}}\nonumber
\end{eqnarray}

Now the problem is to construct such polynomials of any degree but, in particular, to construct a special class of invariant we call fundamental (see below).

Let $\mathbb L$ be the algebra of box diagrams over $\mathbb Q$ and let $\mathbb L^{(n)}$ be the subspace of box diagrams of length $n.$ These are essentially the ordered partitions of $n.$ We can clearly generate all elements of degree $n$ from products of elements of lower degree except for the one new $n$-box element.

The following decomposition is obvious.

\begin{lem}\label{decomposition}
\[{\mathbb L}^n=\bigoplus_{p=1}^n\left(\stackrel{\tiny{\yng(3)}\ldots\tiny{\yng(2)}}{p\,\text{box}}\right).{\mathbb L^{(n-p)}}\]
\end{lem}

The dimension of $\mathbb L^n$ is $2^{n-1}$ \cite{AE}.

Let $\mathbb K$ be the two sided ideal of $\mathbb L$ generated by commutators: 
\[\mathbb K=\{L_A[L_B,L_C]L_D|A,B,C,D\subseteq I\}\]
and let $\mathbb K^{(n)}$ be the subspace of $\mathbb K$ of length $n$ elements. Let $\mathbb S^{(n)}$ be the totally symmetric elements of $\mathbb L^{(n)}.$ Clearly $\mathbb L^{(n)}=\mathbb S^{(n)}\oplus\mathbb K^{(n)}.$

Note that the symmetry here refers to ordering of the components in an expression and is not the same as the symmetry referred to in earlier paragraphs. Thus (writing simply $\mathcal Y$ for $L_{\mathcal Y}$)
\[\tiny{\yng(2).\yng(1)+\yng(1).\yng(2)}\]
is symmetric; 
\[\tiny{\yng(2).\yng(1)-\yng(1).\yng(2)}\] is antisymmetric whereas the corresponding symmetrised, indexed expression
\[\Sigma_{\{i,j,k\}}(\tiny{\yng(2).\yng(1)-\yng(1).\yng(2)})\]
is not zero. (It happens to be a sum of derivatives of second order invariants in this case.)

\begin{thm}
$\delta^{(n)}:\mathbb L^{(n)}\rightarrow\mathbb L^{(n-1)}$ is surjective. Further $\delta^{(n)}:\mathbb S^{(n)}\rightarrow\mathbb S^{(n-1)}$ and $\delta^({n}):\mathbb K^{(n)}\rightarrow\mathbb K^{(n-1)}$ are surjective.
\end{thm}

\emph{Proof.} This follows by induction using the decomposition in lemma (\ref{decomposition}).$\blacksquare$

Since the dimension of $\mathbb S^{(n)}$ is $p(n),$ the number of partitions of $n$, which is a strictly increasing function, $\delta^{(n)}$ has a kernel element in $\mathbb S^{(n)}.$

For instance, in the example above we could equally well have chosen the invariant

\[{\tiny{\yng(3)}}-\frac32{\tiny{\yng(2)}}.{\tiny{\yng(1)}}-\frac32{\tiny{\yng(1)}}.{\tiny{\yng(2)}}+2{\tiny{\yng(1)}}.{\tiny{\yng(1)}}.{\tiny{\yng(1)}}\,.\]

However, we desire a stronger result. We want to show that there exists an invariant in which the leading order term is the $n$-box diagram and all other summands have parts ordered according to decreasing length. We can such an ordering \emph{canonical} and such an invariant \emph{fundamental}. It cannot be an element of $\mathbb S^n.$ There is nevertheless, because of the obvious association with partitions, a bijection between the set of canonically ordered elements in $\mathbb L^n$  and $\mathbb S^n.$  The set of canonically ordered elements of $\mathbb L^n$ is unfortunately not closed under $\delta.$ Most simply, for example,
\[\delta(\tiny{\yng(2).\yng(2)})=\tiny{2\,\yng(1).\yng(2)+2\,\yng(2).\yng(1)}.\]

The map $\delta:\mathbb L^n\rightarrow\mathbb L^{n-1},$ being surjective, has kernel dimension $2^{n-2}.$ Below we will construct a single fundamental invariant at each degree $n$ with leading term the $n$-box diagram having coefficient unity. We call this invariant $I^{(0)}_n.$ It is easy to see that $I_{n-1}^{(1)}=[\tiny{\yng(1)},I^{(0)}_{n-1}],$ $I_{n-2}^{(2)}=[\tiny{\yng(1)},I_{n-2}^{(1)}]$ etc are also invariants of the same order. In fact these objects generate the entire kernel of $\delta.$

\begin{thm}
Given a fundamental invariant $I^{(0)}_n$ at each order $n\geq 2,$ define invariants $I_n^{(p)}=[\tiny{\yng(1)},I_n^{(p-1)}]$ for all $n\geq 2$ and $p>0.$ Then the (non-commutative) kernel of $\delta$ is generated by all products of $I_n^{(p)}$ for $p\geq 0.$
\end{thm}

\emph{Proof.} We will count the number of ordered partitions of length $n$ and show that it is equal to the dimension of the kernel of $\delta_n.$ Because $[\tiny{\yng(1)}, \cdot]$ is a derivative operation it is enough to consider partitions made from the $I_n^{(p)}$ only.

The length of $I_n^{(p)}$ is $n+p$ and $n\geq 2$ so we have $q-1$ objects of length $q\geq 2$ to play with. Let $P_q$ be the number of ordered partitions of length $q.$ Removing the first term gives a partition of the same character but shorter length. Taking account of the number of possible first terms we find (letting $P_0=1$ and $P_1=0$),
\[P_{q}=P_{q-2}+2P_{q-3}+3P_{q-4}+\ldots+(q-3)P_2+(q-1)P_0.\]
Hence
\[P_{q}-P_{q-1}=P_{q-2}+P_{q-3}+P_{q-4}+\ldots + P_2+P_0\]
or
\[P_{q}=P_0+\sum_{i=2}^{q-1}P_{i}.\]

Given $P_2=1$ we get $P_q=2^{q-2}$ which is the dimension of the kernel of $\delta^{(q)}:\mathbb L^{(q)}\rightarrow\mathbb L^{(q-1)}.$ $\blacksquare$

We now give an expression for $I^{(0)}_n.$

\begin{thm}
Let $(r, 1^s)$ denote the (ordered) partition of $n=r+s$ given by $(r,1,1,\ldots 1)$ where there are $s$ repetitions of `1'. We include the case $r=0$ so that $(0,1^n)=(1,1^{n-1}).$ Then an invariant of degree $n$ with leading term $a_I$ is:
\[I^{(0)}_n=\sum_{s=0}^n(-1)^s\binom{n}{s}(n-s,1^s).\]
\end{thm}

\emph{Proof.} This follows by application of $\delta,$
\[\delta(n-s,1^s)=(n-s)(n-s-1,1^s)+s(n-s,1^{s-1})\]
and use of the identity
\[\binom{n}{s}(n-s)=\binom{n}{s+1}(s+1).\quad\blacksquare\]

\section{Examples}\label{eggs}
We write down here the fundamental invariants of degrees up to six.
\subsection*{n=2}
\begin{eqnarray}
&&\tiny{\yng(2)-\yng(1).\yng(1)}\nonumber\\
\\
 I_{2}&=&\overline{L_{12}}-\overline{L_{1}L_{2}}\nonumber\\
           &=&\overline{a_{12}}-\overline{a_{1}a_{2}}-\overline{a_{1,2}}\nonumber
\end{eqnarray}
\subsection*{n=3}
\begin{eqnarray}
&&\tiny{\yng(3)-3\yng(2).\yng(1)+2\yng(1).\yng(1).\yng(1)}\nonumber\\
\\
 I_{3}&=&\overline{L_{123}}-3\overline{L_{12}L_{3}}+2\overline{L_{1}L_{2}L_{3}}\nonumber\\
           &=&\overline{a_{123}}-3\overline{a_{12}a_{3}}+2\overline{a_{1}a_{2}a_{3}}-\overline{a_{1,23}}\nonumber
\end{eqnarray}
\subsection*{n=4}
\begin{eqnarray}
&&\tiny{\yng(4)-4\yng(3).\yng(1)+6\yng(2).\yng(1).\yng(1)-3\yng(1).\yng(1).\yng(1).\yng(1)}\nonumber\\
\\
 I_{4}&=&\overline{L_{1234}}-4\overline{L_{123}L_{4}}+6\overline{L_{12}L_{3}L_{4}}-3\overline{L_{1}L_{2}L_{3}L_{4}}\nonumber\\
           &=&\overline{a_{1234}}-4\overline{a_{123}a_{4}}+6\overline{a_{12}a_{3}a_{4}}-3\overline{a_{1}a_{2}a_{3}a_{4}}-\overline{a_{1,234}}-6\overline{\left(a_{12}-a_{1}a_{2}\right)a_{3,4}}\nonumber\\
&&+3\overline{a_{1,2}a_{3,4}}\nonumber
\end{eqnarray}
\subsection*{n=5}
\begin{eqnarray}
&&\tiny{\yng(5)-5\yng(4).\yng(1)+10\yng(3).\yng(1).\yng(1)-10\yng(2).\yng(1).\yng(1).\yng(1)+4\yng(1).\yng(1).\yng(1).\yng(1).\yng(1)}\nonumber\\
\\
I_{5}&=&\overline{L_{12345}}-5\overline{L_{1234}L_{5}}+10\overline{L_{123}L_{4}L_{5}}-10\overline{L_{12}L_{3}L_{4}L_{5}}+4\overline{L_{1}L_{2}L_{3}L_{4}L_{5}}\nonumber\\
          &=&\overline{a_{12345}}-5\overline{a_{1234}a_{5}}+10\overline{a_{123}a_{4}a_{5}}-10\overline{a_{12}a_{3}a_{4}a_{5}}+4\overline{a_{1}a_{2}a_{3}a_{4}a_{5}}\nonumber\\
          &&-\overline{a_{1,2345}}-10\overline{\left(a_{12}-a_{1}a_{2}-a_{1,2}\right)a_{3,45}}\nonumber\\
          &&-10\overline{\left(a_{123}-3a_{12}a_{3}+2a_{1}a_{2}a_{3}\right)a_{4,5}}\nonumber
\end{eqnarray}
\subsection*{n=6}
\begin{eqnarray}
&&\tiny{\yng(6)-6\yng(5).\yng(1)+15\yng(4).\yng(1).\yng(1)-20\yng(3).\yng(1).\yng(1).\yng(1)+15\yng(2).\yng(1).\yng(1).\yng(1).\yng(1)}\nonumber\\
&&\tiny{-5\yng(1).\yng(1).\yng(1).\yng(1).\yng(1).\yng(1)}\nonumber\\
\\
I_{6}&=&\overline{L_{123456}}-6\overline{L_{12345}L_{6}}+15\overline{L_{1234}L_{5}L_{6}}-20\overline{L_{123}L_{4}L_{5}L_{6}}+15\overline{L_{12}L_{3}L_{4}L_{5}L_{6}}\nonumber\\
&&-5\overline{L_{1}L_{2}L_{3}L_{4}L_{5}L_{6}}\nonumber\\
          &=&\overline{a_{123456}}-6\overline{a_{12345}a_{6}}+15\overline{a_{1234}a_{5}a_{6}}-20\overline{a_{123}a_{4}a_{5}a_{6}}\nonumber\\
&&+15\overline{a_{12}a_{3}a_{4}a_{5}a_{6}}-5\overline{a_{1}a_{2}a_{3}a_{4}a_{5}a_{6}}\nonumber\\
          &&-\overline{a_{1,23456}}-15\overline{\left(a_{12}-a_{1}a_{2}-a_{1,2}\right)a_{3,456}}
          -20\overline{\left(a_{123}-3a_{12}a_{3}+2a_{1}a_{2}a_{3}\right)a_{4,56}}\nonumber\\
&&+10\overline{a_{1,23}a_{4,56}}-15\overline{\left(a_{1234}-4a_{123}a_{4}+6a_{12}a_{3}a_{4}-3a_{1}a_{2}a_{3}a_{4}\right)a_{5,6}}\nonumber\\
&&+45\overline{\left(a_{12}-a_{1}a_{2}\right)a_{3,4}a_{5,6}}-15\overline{a_{1,2}a_{3,4}a_{5,6}}\nonumber
\end{eqnarray}

Thus in the case of a operator $L=\partial_1\partial_2\partial_3\partial_4\partial_5\partial_6+\ldots$ of order six we obtain a single $I_6$ invariant, six $I_5$ invariants, one for each distinct choice of five indices from six, ten $I_4$ invariants and so on, giving us a total of $2^6-6-1=57$ fundamental invariants.

In general we will have for the $n^{th}$ order differential operator
\[\sum_{s=2}^n\binom{n}{s}=2^n-n-1\]
such invariants.

\section{Conclusion}

We have presented generalizations of Laplace invariants for scalar, linear, hyperbolic partial differential equations of a reasonably general form and of arbitrary order. This amounts to construction of the central part of a ring of differential operators simply related to the given one and the invariants are expressed as linear sums of ordered partitions of integers. The list of inavariants given is not complete in that it omits those not symmetric in indices. 

Clearly the next important step is to formulate all possible Laplace, Darboux and Moutard transformations for these invariants as well as to relate them to results on systems. It would be pleasing to adopt the philosophy of this paper and present, say, Laplace transformations as in a purely operator based fashion.

One may also ask what general relations between invariants will allow for factorization.


\begin{thebibliography}{99}
\bibitem{AE} Andrews, G.E and and Eriksson, K., \emph{Integer Partitions}, CUP (2004).
\bibitem{A} Athorne, C., \emph{A $\mathbb Z^2\times\mathbb R^3$ Toda System}, Physics Letters A, 206 (1995) 162--166.
\bibitem{AY} Athorne, C. and Yilmaz, H.,\emph{Laplace Invariants for General Hyperbolic Systems}, Journal of Nonlinear Mathematical Physics, Vol. 19, No. 3 (2012) 391--410.
\bibitem{BKar} Beals, R. and Kartashova, E.A., \emph{Constructively factoring linear partial differential operators in two variables}, Theoretical and Mathematical Physics, 145 (2005) 1511--1524.
\bibitem{C} Coutinho, S.C., \emph{A Primer of Algebraic D-Modules}, LMS Student Text 33 (1995).
\bibitem{D} Darboux, G., \emph{Le\c{c}ons Sur La Th\'eorie G\'en\'erale Des Surfaces Et Les Applications G\'eom\'etriques Du Calcul Infinit\'esimal}, Gauthier-Villars (1887-96).
\bibitem{Dz} Dzhokhadze, O.M., \emph{Laplace Invariants for some classes of linear partial differential equation}, Differ. Uravn. 40 (2004) 58--68, 142; translation in
Differ. Equ. 40 (2004) 63--74.
\bibitem{F} Ferapontov, E. V., \emph{Laplace transforms of hydrodynamic-type systems in Riemann invariants}, Teoret. Mat. Fiz. 110 (1997) 86--97; translation in Theoret. and Math. Phys. 110 (1997) 68--77.
\bibitem{G} Gousat, E., \emph{Sur une transformation de l'\'equation $s^2=4\lambda(x,y)pq$}, Bull. de la Soc. Math. de France, 28 (1900) 1--6.
\bibitem{GZ} Guryeva, A.M. and Zhiber, A.V., \emph{Laplace invariants of two-dimensional open Toda lattices}, Theoretical and Math. Phys., 138 (2004) 338--355.
\bibitem{JA} Jur\'a\v{s}, M. and Anderson, I.M., \emph{Generalized Laplace invariants and the method of Darboux}, Duke Math. J. 89 (1997) 351--375.
\bibitem{Kam} Kamran, N., \emph{Selected topics in the geometrical study of differential equations}, CBMS Regional Conference Series in Mathematics, 96, AMS, Providence, RI, 2002.
\bibitem{KamT} Kamran, N. and Tenenblat, K., \emph{Laplace transformations in higher dimensions}, Duke Math. J. 84 (1996) 237--266.
\bibitem{Kar} Kartashova, E.A., \emph{A hierarchy of generalized invariants for linear partial differential operators}, Theroretical and Mathematical Physics, 147 (2006) 839-846.
\bibitem{M} Mironov, A.N., \emph{On the Laplace Invariants of a Fourth-Order Equation}, Differential Equations, Vol. 45, No. 8, 1168--1173 (2009). 
\bibitem{MM} Mironov, A.N. and Mironova, L.B., \emph{Laplace invariants for a fourth order equation with two independent variables}, Izv. Vyssh. Uchebn. Zaved. Mat., 2014, no. 10, 27--34.
\bibitem{Md} Moutard, Th F., \emph{Sur la construction des équations de la forme $\frac{\partial^2}{\partial_x\partial_y}= \lambda(x, y)$ qui admettenent une int\'egrale g\'en\'erale explicite}, J. Ec. Pol 45 (1878): 1--11.
\bibitem{S} Shemyakova, E., \emph{A Full System of Invariants for Third-Order Linear Partial Differential Operators}, In: Lecture Notes in Computer Science 4120 (2006), J.  Calmet, T. Ida, D. Wang (Eds.), Springer.
\bibitem{SE} Shemyakova, E. and Mansfield, E.L., \emph{Moving frames for Laplace invariants}, ISSAC 2008, 295-302, ACM, New York, 2008.
\bibitem{SW} Shemyakova, E. and Winkler, F., \emph{Obstacles to the Factorization of Linear Partial Differential Operators into Several Factors}, Programming and Computer Software, vol.33, no.2, pp.67--73, 2007
\bibitem{T} Toda, M., \emph{Theory of nonlinear lattices}, Springer (1989).
\bibitem{Ts1} Tsar\"ev, S.P., \emph{Factorization of linear differential operators and systems}, in \emph{Algebraic Theory of Differential Equations} ed. Maccallum, A.H. and Mikhailov, A. V., LMS lecture note series 357, CUP (2009).
\bibitem{W} Weiss, J., \emph{B\"acklund transformations, focal surfaces and the two-dimensional Toda lattice}, Physics Letters A 137 (1989) 365--368. 
\bibitem{Ts2} Tsar\"ev, S.P., \emph{Generalized Laplace transformations and integration of hyperbolic systems of linear partial differential equations}, ISSAC'05, 325--331, ACM, New York, 2005. 
\bibitem{ZS} Zhiber, A.V. and Sokolov, V.V., \emph{Exactly integrable hyperbolic equations of Liouville type}, Uspekhi mat. nauk, 2001, 56, no.1, 63-106   [in Russian]; translation in Russian Math. Surveys, 2001, 56, no.1, 61–101
\end{thebibliography}
\end{document}